\documentstyle [11pt,a4]{article}
\newcommand{\De}{\Delta}

\newcommand{\la}{\lambda}

\newcommand{\Lao}{\Lambda_0}
\newcommand{\si}{\sigma}

\newcommand{\vep}{\varepsilon}

\newcommand{\vk}{\vec{k}}
\newcommand{\vp}{\vec{p}}
\newcommand{\vq}{\vec{q}}
\newcommand{\vph}{\varphi}

\newcommand{\ti}[1]{\tilde{#1}}
\newcommand{\qed}{\hfill \rule {1ex}{1ex}\\ }

\newcommand{\eq}{\begin{equation}}
\newcommand{\eqe}{\end{equation}}
\font \smallescriptscriptfont = cmr5
\font \smallescriptfont       = cmr5 at 7pt
\font \smalletextfont         = cmr5 at 10pt
\font \tensans                = cmss10
\font \fivesans               = cmss10 at 5pt

\font \sevensans              = cmss10 at 7pt

\newfam\sansfam
\textfont\sansfam=\tensans\scriptfont\sansfam=\sevensans
\scriptscriptfont\sansfam=\fivesans
\def\sans{\fam\sansfam\tensans}
\def\bbbr{{\rm I\!R}} 

\def\bbbc{{\mathchoice {\setbox0=\hbox{$\displaystyle\rm C$}\hbox{\hbox
to0pt{\kern0.4\wd0\vrule height0.9\ht0\hss}\box0}}
{\setbox0=\hbox{$\textstyle\rm C$}\hbox{\hbox
to0pt{\kern0.4\wd0\vrule height0.9\ht0\hss}\box0}}
{\setbox0=\hbox{$\scriptstyle\rm C$}\hbox{\hbox
to0pt{\kern0.4\wd0\vrule height0.9\ht0\hss}\box0}}
{\setbox0=\hbox{$\scriptscriptstyle\rm C$}\hbox{\hbox
to0pt{\kern0.4\wd0\vrule height0.9\ht0\hss}\box0}}}}

\def\bbbe{{\mathchoice {\setbox0=\hbox{\smalletextfont e}\hbox{\raise
0.1\ht0\hbox to0pt{\kern0.4\wd0\vrule width0.3pt height0.7\ht0\hss}\box0}}
{\setbox0=\hbox{\smalletextfont e}\hbox{\raise
0.1\ht0\hbox to0pt{\kern0.4\wd0\vrule width0.3pt height0.7\ht0\hss}\box0}}
{\setbox0=\hbox{\smallescriptfont e}\hbox{\raise
0.1\ht0\hbox to0pt{\kern0.5\wd0\vrule width0.2pt height0.7\ht0\hss}\box0}}
{\setbox0=\hbox{\smallescriptscriptfont e}\hbox{\raise
0.1\ht0\hbox to0pt{\kern0.4\wd0\vrule width0.2pt height0.7\ht0\hss}\box0}}}}

\def\bbbq{{\mathchoice {\setbox0=\hbox{$\displaystyle\rm Q$}\hbox{\raise
0.15\ht0\hbox to0pt{\kern0.4\wd0\vrule height0.8\ht0\hss}\box0}}
{\setbox0=\hbox{$\textstyle\rm Q$}\hbox{\raise
0.15\ht0\hbox to0pt{\kern0.4\wd0\vrule height0.8\ht0\hss}\box0}}
{\setbox0=\hbox{$\scriptstyle\rm Q$}\hbox{\raise
0.15\ht0\hbox to0pt{\kern0.4\wd0\vrule height0.7\ht0\hss}\box0}}
{\setbox0=\hbox{$\scriptscriptstyle\rm Q$}\hbox{\raise
0.15\ht0\hbox to0pt{\kern0.4\wd0\vrule height0.7\ht0\hss}\box0}}}}

\def\bbbt{{\mathchoice {\setbox0=\hbox{$\displaystyle\rm
T$}\hbox{\hbox to0pt{\kern0.3\wd0\vrule height0.9\ht0\hss}\box0}}
{\setbox0=\hbox{$\textstyle\rm T$}\hbox{\hbox
to0pt{\kern0.3\wd0\vrule height0.9\ht0\hss}\box0}}
{\setbox0=\hbox{$\scriptstyle\rm T$}\hbox{\hbox
to0pt{\kern0.3\wd0\vrule height0.9\ht0\hss}\box0}}
{\setbox0=\hbox{$\scriptscriptstyle\rm T$}\hbox{\hbox
to0pt{\kern0.3\wd0\vrule height0.9\ht0\hss}\box0}}}}

\def\bbbs{{\mathchoice
{\setbox0=\hbox{$\displaystyle     \rm S$}\hbox{\raise0.5\ht0\hbox
to0pt{\kern0.35\wd0\vrule height0.45\ht0\hss}\hbox
to0pt{\kern0.55\wd0\vrule height0.5\ht0\hss}\box0}}
{\setbox0=\hbox{$\textstyle        \rm S$}\hbox{\raise0.5\ht0\hbox
to0pt{\kern0.35\wd0\vrule height0.45\ht0\hss}\hbox
to0pt{\kern0.55\wd0\vrule height0.5\ht0\hss}\box0}}
{\setbox0=\hbox{$\scriptstyle      \rm S$}\hbox{\raise0.5\ht0\hbox
to0pt{\kern0.35\wd0\vrule height0.45\ht0\hss}\raise0.05\ht0\hbox
to0pt{\kern0.5\wd0\vrule height0.45\ht0\hss}\box0}}
{\setbox0=\hbox{$\scriptscriptstyle\rm S$}\hbox{\raise0.5\ht0\hbox
to0pt{\kern0.4\wd0\vrule height0.45\ht0\hss}\raise0.05\ht0\hbox
to0pt{\kern0.55\wd0\vrule height0.45\ht0\hss}\box0}}}}

\def\bbbz{{\mathchoice {\hbox{$\sans\textstyle Z\kern-0.4em Z$}}
{\hbox{$\sans\textstyle Z\kern-0.4em Z$}}
{\hbox{$\sans\scriptstyle Z\kern-0.3em Z$}}
{\hbox{$\sans\scriptscriptstyle Z\kern-0.2em Z$}}}}
\begin{document}
\title{ Singularity Cancellation in Fermion Loops through Ward
  Identities}
\maketitle

\centerline{C. Kopper and J. Magnen}

\centerline{Centre de Physique Th{\'e}orique, CNRS UPR 14}
\centerline{Ecole Polytechnique} \centerline{91128 Palaiseau Cedex,
  FRANCE} \vskip 1cm \medskip
\noindent
{\bf \centerline{Abstract}}\\
Recently Neumayr and Metzner [1] have shown that the connected $N$-point 
density-correlation functions of the two-dimensional 
and the one-dimensional Fermi gas at one-loop order    
generically (i.e. for nonexceptional
energy-momentum configurations) 
vanish/are regular in the small momentum/small energy-momentum limits.
Their result is based on an explicit analysis in the sequel of the
results of Feldman et al. [2].
In this note we use Ward identities to give a proof 
of the same fact  - in a considerably shortened and
simplified way - for any dimension of space.\newpage 

The infrared properties of  the connected $N$-point 
density-correlation function of the interacting Fermi 
gas at one-loop order, to be called $N$-loop for shortness, 
are  important for the understanding 
of interacting Fermi systems, in particular in the low
energy regime. The $N$-loops appear as Feynman (sub)diagrams or as kernels 
in effective actions. In two dimensions e.g.,  their properties are relevant 
for the analysis of the electron gas
in relation with questions such as the breakdown of Fermi liquid
theory and high temperature superconductivity.      
We refer to the literature in this respect, see [3,4] and references given
there. Whereas  the contribution of a single loop-diagram to the
$N$-point function for $N\ge 3\,$ 
generally diverges in the small energy-momentum
limit, these singularities have been known to cancel each other in
various situations [3,4,5] in the symmetrized contribution, i.e. when
summing over all possible orderings of the external momenta,
a phenomenon called loop-cancellation.
The two-loop has been known
explicitly in one, two and three dimensions  for quite some time [1], 
the calculation in two dimensions goes back to Stern [6].  
We introduce the following notations adapted to those of [1]~:\\
$\Pi_N(q_1,\ldots, q_{N})\,$ denotes the Fermionic $N$-loop
for $N\ge 3\,$,  see (\ref{intt}) below,
as a function of the (outgoing) external energy-momentum variables 
$q_1,\,q_2\ldots, q_{N-1}\,$ and $\,q_N=-(q_1+\ldots+ q_{N-1})\,$. 
Here the $(d+1)$-vector $q\,$ stands for 
$\,(q_0,\,q_1,\ldots,q_d)=(q_0,\vq\,)\,$.
We also introduce the variables 
\eq
p_i\,=\,q_1\,+\,q_2\,+\,\ldots +\,q_{i-1}\,,\quad p_1\,=\,0\,,
\quad 1 \le i \le N\ .
\eqe
By definition we then have
\eq
\Pi_N(q_1,\ldots, q_{N})=
\int\!\! {dk_0 \over 2\pi}\,
{d^dk \over (2\pi)^d}\,
I_N(k;\,q_1,\ldots,q_N)\
\mbox{ with }\ 
I_N(k;\,q_1,\ldots,q_N)
=\,\prod_{j=1}^N  G_0(k-p_j) 
\label{intt}
\eqe
\[\mbox{ and }
\,\,G_0(k)= {1 \over ik_0 -(\vep_{\vk}-\mu)}\,,\ 
\,\vep_{\vk}= {\vk^2\over 2m}\,, 
\ \mu\, \mbox{ being the Fermi energy.}
\]
To have absolutely convergent integrals for $\,N\ge 3\,$,
we restrict the subsequent considerations to the physically
interesting cases $\,d\le 3\,$. At the end of the paper
we indicate how the same results can be obtained
for  $\,d\ge 4\,$. We also assume that the variables $\,q_j\,$
have been chosen such that the integrand is not singular (see below 
(\ref{nonex})). 
In the following we will choose units such that $\mu=1,\ 2m=1\,$.
By convention the vertex of $q_1\,$ 
will be viewed as  the first vertex.\\ 
Symmetrization with respect to the external
momenta $(q_1,\ldots, q_{N})\,$  diminishes the degree of
singularity of the Fermion loops. To prove this fact we have to introduce 
some notation on permutations. We denote by $\si\,$ any permutation
of the sequence $(2,\ldots, N)\,$.
By
$\Pi_N^{\si}(q_1,\ldots, q_{N})\,$ we then denote
$\Pi_N(q_1,q_{\si^{-1}(2)},\ldots, q_{\si^{-1}(N)})\,$.
For the completely symmetrized $N$-loop
we write\footnote{We do not divide 
by the number of permutations, here  $\,(N-1)!\,$, to shorten 
some of the subsequent formulae.}
\eq
\Pi_N^S(q_1,\ldots,q_{N})\,=\,
\sum_{\si} \Pi_N^{\si}(q_1,\ldots, q_{N})\ .
\label{symm}
\eqe
We will also have to consider subsets of permutations~:
For $n \le N-2\,$ and $\,2 \le j_1<j_2<\ldots <j_n\le N\,$ we denote by
$\,\si^{(i_1,\ldots,i_n)}_{(j_1,\ldots, j_n)}\;$
the permutation mapping $j_{\nu} \to i_{\nu}=\si(j_{\nu}) 
\in \{2,\ldots,N\}\,$, which preserves the order of the 
remaining sequence $\,\Bigl( (2,\ldots,N)-(j_1,\ldots, j_n) \Bigr)\,$,
i.e. $\,\si(\nu) < \si(\mu)\,$ for $\nu < \mu\,$, if
$\,\nu,\, \mu \not\in \{j_1,\ldots, j_n\}\,$. 
When the target positions $\,(i_1,\ldots,i_n)\,$
are summed over (see e.g. (\ref{symn}) below),  
we will write shortly $\si(j_1,\ldots, j_n)\,$, 
or  also $\si_N(j_1,\ldots, j_n)\,$,  $\,\si_N^n\,$, 
if we want to indicate the number $\,N\,$.
Note that $n = N-2\,$ is already the most general case,
since fixing the positions of $N-2\,$ variables (apart from $q_1\,$)
fixes automatically that of the last.
For the permutation $\,\si_{(j)}^{(i)}\,$,
which maps $j\,$ onto the 
$i$-th position in the sequence $(2,\ldots,N)\,$ (preserving the order
of the other variables), we use the shorthands $\si^i_{j}\,$ 
or $\si_{j}\,$.
We then also introduce the $N$-loop, 
symmetrized with respect to the previously 
introduced subsets of permutations, i.e.\footnote{Again we do not
multiply by  ${(N-n-1)! \over (N-1)!}\,$.}
\eq
\Pi_N^{S_n(j_1,\ldots, j_n)}(q_1,\ldots,q_{N})=\!\!\!
\sum_{\si_N(j_1,\ldots, j_n)} 
\Pi_N^{\si_N(j_1,\ldots, j_n)}(q_1,\ldots, q_{N})\,,
\ \mbox{ in particular } \ \Pi_N^S\,=\,\Pi_N^{S_{N-2}}\ .
\label{symn}
\eqe
The  notations corresponding  to 
(\ref{symm} - \ref{symn}) will be applied in the same sense 
also to $I_N\,$. 

The recent result [1] of Neumayr and Metzner, based on the exact
expression for the $N$-loop from [2], 
which however is nontrivial to analyse, 
shows that for $N>2\,$  and $d=1,\,2\, $ one has generically ~:\\    
\eq
\Pi_N^S(\la q_1,\ldots,\la q_{N})=\,O(1)\
\mbox{ for } \ \la\to 0  \;,
\label{me1}
\eqe
\eq
\Pi_N^S(q_{10}\,,\,\la \vq_1\,,\ldots,\,q_{N0}\, 
,\,\la \vq_{N})=\,O(\la ^{2N-2})\ 
\mbox{ for } \ \la\to 0 \;,
\label{me2}
\eqe
\eq
\Pi_N^S(q_1,\ldots, q_{N})=\,O(|\vq_j|)\
\mbox{ for } \ \vq_j\to 0 \ .
\label{me3}
\eqe
We are not completely sure about the authors'
definition of 'generically'. In any case
their restrictions on the energy-momentum 
variables include the following one:\\
The energy momentum set $\, \{q_1,\ldots,q_N\}\,$ is {\it nonexceptional},
 if for all $\,J\subset_{\neq} \{1,\ldots,N\}\,$ we have 
\eq
|\sum_{i\in J} q_{i0}| \ge \eta > 0\ .
\label{nonex}
\eqe
Our bounds given in the subsequent proposition are based on this
condition.\footnote{As for the spatial components $\vq_i\,$
we only suppose that they lie in some fixed compact region 
$\{|\vq_j|\le K\,\}$.}. 
Though we cannot exclude (and did not
really try) that linear relations among the momentum variables
$\,\{\vq_1,\ldots,\vq_N\}\,$ could even improve those bounds, 
it seems quite clear that they are saturated apart from
subsets of momentum configurations of measure zero, cf. also the
numerical results mentioned in [1]. Furthermore they
deteriorate with the parameter $\eta^{-1}\,$
(cf. the remarks in the end of the paper).\\[.2cm]
\noindent
{\bf Proposition}~: {\sl For nonexceptional energy-momentum
configurations $\{q_{1},\ldots, q_{N}\}\,$ 
(as defined through (\ref{nonex}) with $\eta\,$ fixed)
and for $\,N\ge 3\,$ and $n \le N-2\,$ the following bounds hold: 
\eq
A)\quad\qquad \qquad \quad\mbox{ In the small }\, \la\, \mbox{ limit }  
\ q_{i0}\to \la \,q_{i0}\,,\
\vq_i \to \la  \,\vq_i \,,\
\la \to \,0 \qquad\quad
\label{sc}
\eqe 
\eq
A1)\qquad\qquad\qquad
|\Pi_N(\la q_1,\ldots,\la q_{N})|\le\,O(\la ^{-(N-2)})\,\mbox{,\hspace{4cm}}
\label{pc}
\eqe
\eq
A2)\quad \ \, 
|\Pi_N^{S_n}(\la q_1,\ldots,\la q_{N})|\le\,O(\la^{-(N-2-n)})\,,
\quad |\Pi_N^S(\la q_1,\ldots,\la q_{N})|\le\,O(1)\ .
\label{pcsy}
\eqe
\eq
B) \qquad \qquad\qquad \mbox{ In  the dynamical limit }
\,\vq_i \to \la  \vq_i \,,\
\la \to \,0\qquad\qquad\qquad\qquad
\label{dynl}
\eqe 
\eq
 B1) \qquad\qquad\qquad 
|\Pi_N(q_{10}\,,\,\la \vq_1\,,\ldots,\,q_{N0}\, 
,\,\la \vq_{N})|\,\le\,O(\la^2)\;,\qquad\qquad\qquad\qquad
\label{pcimp}
\eqe
\eq
 B2)\quad \,\,
\,|\Pi_N^{S_n}(q_{10}\,,\,\la \vq_1\,,\ldots)|\le\,O(\la ^{2n+2})\,,\quad 
\;|\Pi_N^S(q_{10}\,,\,\la \vq_1\,,\ldots)|\le\,O(\la
^{2N-2}) \ .
\label{pcimpsy}
\eqe
The functions $\,\la^{N-2}\,
\Pi_N(\la q_1,\ldots,\la q_{N})\,$,
$\,\Pi_N^S(\la q_1,\ldots,\la q_{N})\,$,
$\,\Pi_N(q_{10}\,,\,\la \vq_1\,,\ldots,\,q_{N0},\,\la \vq_{N})\,$
and\\ $\,\Pi_N^S(q_{10}\,,\,\la \vq_1\,,\ldots,\,q_{N0}\, 
,\,\la \vq_{N})\,$ are analytic functions
of $\,\la\,$ in a neighbourhood of $\la=0\,$ (depending on the
momentum configuration, in particular on $\eta\,$).}\\[.2cm]
{\sl Proof~:}  
To prove A1) we perform the
$\,k_0$-integration using residue calculus, so that (\ref{intt}) 
takes the form (cf. [2,1])
\eq
\Pi_N(q_1,\ldots, q_{N})\,=\,
\sum_{i=1}^N \int_{|\vk-\vp_i| < 1} {d^dk \over (2\pi)^d}\,
\,\,\Bigl(\prod_{j=1,j\neq i}^N \,\,f_{ij}(\vk)\Bigr)^{-1}\ , 
\mbox{ where }
\label{int1}
\eqe
\[
f_{ij}(\vk)= \vep(\vk-\vp_i)-\vep(\vk-\vp_j)+i  
(p_{i0}- p_{j0})\,=\,
2\vk\cdot(\vp_j-\vp_i)+ (\vp_i^{\,2}-\vp_j^{\,2})+i  
(p_{i0}- p_{j0})\ .
\]
This implies that
\eq
\Pi_N(\la q_1,\ldots, \la q_{N})\,=\,\la^{-(N-1)}
\sum_{i=1}^N \int_{|\vk-\la \vp_i| < 1} {d^dk \over (2\pi)^d}\,
\,\,\Bigl(\prod_{j=1,j\neq i}^N \,\,f^{\la}_{ij}(\vk)\Bigr)^{-1}\ , 
\mbox{ where }
\label{intla}
\eqe
\[
f^{\la}_{ij}(\vk)\,=\,
2\vk\cdot(\vp_j-\vp_i)+ \la (\vp_i^{\,2}-\vp_j^{\,2})+i  
(p_{i0}- p_{j0})\ .
\]
By Lemma 1 below we find
\eq
\sum_{i=1}^N \int_{|\vk| < 1} {d^dk \over (2\pi)^d}\,
\,\,\Bigl(\prod_{j=1,j\neq i}^N \,\,f^{\la}_{ij}(\vk)\Bigr)^{-1} =\,0
\label{int0}
\eqe
so that we obtain 
\eq
\Pi_N(\la q_1,\ldots, \la q_{N})\,=\,
\,\la^{-(N-1)}
\sum_{i=1}^N \{\int_{|\vk-\la \vp_i| < 1}\!\!-\int_{|\vk| < 1}\,\} 
{d^dk \over (2\pi)^d}\,\,
\Bigl(\prod_{j=1,j\neq i}^N \,\,f^{\la}_{ij}(\vk)\Bigr)^{-1}\ .
\label{intla2}
\eqe
Thus each entry in the sum in (\ref{intla}) has to be integrated
only over a domain of measure $O(\la)\,$. 
Since the integrands are bounded in modulus  by $\,O(1)\,$ due to the
nonexceptionality of the momenta,
this leads to the statement (\ref{pc}).\\[.2cm]
To prove B1), (\ref{pcimp}) we use again
(\ref{int1})
\eq
\Pi_N(q_{10}\,,\,\la \vq_1\,,\ldots,\,q_{N0}\, 
,\,\la \vq_{N})\,=\,
\sum_{i=1}^N\{\int_{|\vk-\la \vp_i| < 1}\!\!-\int_{|\vk| < 1}\,\}
 {d^dk \over (2\pi)^d}\,
\,\,\Bigl(\prod_{j=1,j\neq i}^N \,\,f_{ij}(\la;\,\vk)\Bigr)^{-1}
\label{intlaa}
\eqe
\[
\mbox{ with }\quad f_{ij}(\la;\,\vk)\, =\,
2\la \,\vk\cdot(\vp_j-\vp_i)+ \la^2\,(\vp_i^{\,2}-\vp_j^{\,2})+i  
(p_{i0}- p_{j0})\ .
\]
On performing the change of variables
$\,\ti{\vk}=\,\vk -\la \vp_i\,$
in the  integral over $|\vk-\la \vp_i| < 1\,$ one realizes
that the difference between the two integrals is of order
$\la^2\,$. Or one may convince oneself that both  integrals
are even functions of $\la\,$. In any case this proves B1).
\\[.2cm]
The previous considerations also imply that
$\,
\la^{N-2}\;\Pi_N(\la q_1,\ldots, \la q_{N})
\,\,\mbox{ and }
\,\,\Pi_N(q_{10}\,,\,\la \vq_1\,,\ldots)\,
\,$
are analytic around $\,\la=0\,$:
The imaginary parts of the deno\-minators in (\ref{int1},
\ref{intla}) stay bounded away from zero, and a convergent Taylor
expansion for $\,\la^{N-2}\times$(\ref{intla2}) is easily 
obtained on performing the change of variables  
$\,\ti{\vk}=\,\vk -\la \vp_i\,$
in the first integrals.\\
For the proof of the proposition
we will need also a slight generalization of the bounds on
(\ref{int1}), which  we have just obtained. 
We have to regard integrals of the
type
\eq
\Pi_N(\vph;\,q_1,\ldots, q_{N})~:=
\int\!\! {dk_0 \over 2\pi}\,
{d^dk \over (2\pi)^d}\,
I_N(k;\,q_1,\ldots,q_N)\ \vph(\vk;\,q_1,\ldots,q_N)\ ,
\label{intvp}
\eqe
where we demand that the functions $\,\vph(\vk;\,q_1,\ldots,q_N)\,$
be continuous and uniformly bounded in the domain specified by
(\ref{nonex})~:
$\ |\vph(\vk;\,q_1,\ldots,q_N)| \le C\,$  for  some  
suitable   $\,C>0\,$, and furthermore  
\eq
|\vph(\vk;\,q_1,\ldots,q_N) \,-\,\vph(\vk+\vq;\,q_1,\ldots,q_N) | 
\le \,\sup_J\, |\vq\cdot {\vq}_J| \;C \quad
\mbox{uniformly in }\  \vq
\in  \bbbr^d \ . 
\label{bd}
\eqe
Here we set
\eq
{\vq}_J =\sum_{j\in J}\vq_j
\quad \mbox{for } J \subset \{1,\ldots,N\}\ .
\eqe
The scaling and dynamical
limits A1) and B1) can also be studied 
for $\,\Pi_N(\vph;\,q_1,\ldots, q_{N})\,$
on introducing
\eq
\vph_s(\vk;\,q_1,\ldots, q_{N})=
\vph(\vk;\,\la q_1,\ldots, \la q_{N})\,,  \
\vph_d(\vk;\,q_1,\ldots, q_{N})=
\vph(\vk;\,q_{10},\la \vq_1,\ldots, \,q_{N0},\la \vq_N)\; .
\label{sd}
\eqe
To perform the $\,k_0\,$-integration as before, it is important
to note that $\,\vph\,$ does not depend on  $\,k_0\,$.
The bounds 
\eq
|\Pi_N(\vph_s;\,\la q_1,\ldots, \la q_{N})|\,\le 
O(\la ^{-(N-2)})\,,\quad
|\Pi_N(\vph_d;\,q_{10},\la \vq_1,\ldots, \,q_{N0},\la \vq_N)|\,\le 
O(\la^{2}) 
\label{int22}
\eqe
are then proven as previously. In particular to prove (\ref{pcimp})
we perform the same change of variables as after  (\ref{intlaa})
and use (after telescoping the integrand)
\[
|\vph_d(\ti{\vk}+\la \vp_i;\,q_1,\ldots,q_N )\,
-\,\vph_d(\ti{\vk};\,q_1,\ldots,q_N )|\,\le\,
O(\la ^2)
\]
for $|\vp_i| \le K\,$,
to obtain the factor of $\,\la ^2\,$ required in B1).
We finally note that (\ref{int22}) also holds
for $\,N\ge 2\,$, by same method of proof, if the function $\vph$
assures the integrability of the integrand. On multiplying
any admissible $\vph\,$
by the function $\,\De\,$ or 
from (\ref{dede}),
this is assured, and $\,\De\,\vph \,$ 
has the properties required for
$\vph\,$ above. Therefore we also find for $N\ge 2\,$
\eq\!
|\Pi_N((\De \,\vph)_s;\,\la q_1,\ldots, \la q_{N})|\le 
O(\la^{-(N-2)})\,,\
|\Pi_N((\De\,\vph)_d;\,q_{10},\la \vq_1,\ldots, \,q_{N0},\la \vq_N)|\le 
O(\la^{2})\, ,\!
\label{int23}
\eqe
and the same bound for $\,\Pi_N({1 \over A}\, \vph)\,$,
using $\,{1 \over A}\,$  from (\ref{ad}) (in case 
$\,\vq_j\,$ does not vanish).
 
\noindent
It remains to prove\\[.2cm]  
{\bf Lemma 1~:} For any $n \ge 2 \,$ and pairwise distinct
complex numbers  
$\,a_i\,$, $i \in \{1,\ldots,n\}\,$,
we set $\,a_{i,j}\,= a_i-a_j\, $. Then we have
\eq
\sum^{n}_{i=1} \prod_{j\neq i,j=1}^n {1 \over a_i-a_j\,  }\,=\,0\ .
\label{sum0}
\eqe
{\sl Proof~:} 
By isolating the term $\, i=1\,$ in the sum 
\eq
\sum^{n}_{i=1} \prod_{j\neq i,j=1}^{n}{1 \over a_i-a_j\, }
=
\prod_{j=2}^{n}{1 \over a_1-a_j\, }\,-\,
\sum^{n}_{i=2} \bigl(\prod_{j\neq i,j=2}^{n}{1 \over a_i-a_j\, }\bigr)
{1 \over a_1-a_i}
\label{da}
\eqe
we obtain a presentation of (\ref{sum0}) in terms of a difference 
of two rational functions of the complex variable $\,a_1\,$. They both
have simple poles at $\,a_2,\,\ldots\,,a_n\,$ with identical residues.
So the left hand side is an entire function of  $\,a_1\,$ 
which  vanishes for $\,|a_1| \to \infty\,$ and thus equals zero. 
(This implies that the second
term on the right hand side is the partial fraction expansion of the
first.)
\qed
\\[.2cm]
We now want to show how to obtain the statements A2) and B2) 
from A1) and B1) using the 
{\it Ward identity},\footnote{When regarding more general situations,
  a more general form of
this identity can be derived from the functional integral defining
the interacting fermion theory, in a way analogous to the famous  Ward
identity  of QED. This identity between $\,N$- and 
$N\!-\!1$-point functions is related to fermion number
conservation. In the present case we avoid introducing functional
integrals and restrict to the simple propagator identity (\ref{WIprop}).} 
in form of the simple propagator identity
\eq
(iq_0-2\vq \cdot \vk +\vq^{\,2})\, G_0(k-q)\,G_0(k)\,=\,
G_0(k-q)\,-\,G_0(k)\ .
\label{WIprop}
\eqe
When applying this identity to the
product of the two subsequent propagators in $\,I_N\,$
(\ref{intt}), which differ  by the momentum  $q_j\,$, and then 
summing over all possible positions of the momentum
$q_j\,$ in the loop, the sum telescopes, and  
we are left with the very first and very last contributions, the last 
being obtained  from the first on shifting the variable $\,k\,$ 
by $q_j\,$. Thus we obtain for $\,j \in \{2,\ldots,N\}\,$
\eq
\sum_{\si_{j}} 
 \{iq_{j,0}- 2 \vq_j \cdot
(\vk\,-\,\vp^{\,\si_{j}})+\vq^{\,2}_j\}\,   
I_N^{\si_j}(k;\,q_1, q_2,\ldots, q_{N})\,=\,
\label{WI}
\eqe 
\[
I_{N-1}(k;\,q_1+q_j, q_2, \ldots,\not\!  q_j, \ldots, q_{N})\,-\,
I_{N-1}(k+q_j;\,q_1+q_j, q_2, \ldots,\not\! q_j, \ldots, q_{N})\ .
\]
Note that the term on the r.h.s. vanishes on integration over $k\,$.
The momentum
$p^{\si_{j}}\,$, short for  $p^{\si_{j}^i}\,$, 
is defined to be the momentum arriving at the
vertex of $q_j\,$ for the permutation $\si_{j}^i\,$, i.e.
\eq
p^{\si^i_j}=q_1+\,q_{\si^{-1}(2)}+ \ldots +q_{\si^{-1}(i-1)}=
\left \{\!\! \begin{array}{r@{,}l} 
\,q_1+\ldots+q_{j-1}+q_{j+1}+\ldots +q_{i}
  & \ \mbox{ for }
\quad i>j  \\  
q_1+\ldots+q_{i-1}
 & \ \mbox{ for }
\quad i<j\,.  \\
\end{array}  \right\} 
\label{si}
\eqe
We can rewrite (\ref{WI}) as 
\eq
A(\vk,q_j)\,I_N^{S_1(j)}(k;\,q_1,\ldots, q_{N})\,=\,
-\,\sum_{\si_{j}} 2 \vq_j \cdot\vp^{\,\si_{j}}
I_N^{\si_j}(k;\,q_1,\ldots, q_{N})\,+\,
\label{WI1}
\eqe 
\[
I_{N-1}(k;\,q_1+q_j,\ldots,\not\!  q_j, \ldots, q_{N})\,-\,
I_{N-1}(k+q_j;\,q_1+q_j,\ldots,\not\! q_j, \ldots, q_{N})
\]
\eq
\mbox{ with the definition }\qquad
A(\vk,q_j)\,=\,
 iq_{j0}- 2 \vq_j \cdot
\vk \,+\,\vq^{\,2}_j\ .   
\label{ad}
\eqe
On dividing by $\,A(\vk,q_j)\,$, which is bounded away from $\,0\,$ 
due to (\ref{nonex}), we obtain (in shortened notation)
\eq
I_N^{S_1(j)}(k;\,q_1,\ldots)=\,
-\,\sum_{\si_{j}} {2 \over A(\vk,q_j)}\, \vq_j \cdot\vp^{\,\si_{j}}
I_N^{\si_j}(k;\,q_1,\ldots)\,+\;
\De(\vk,q_j)\,I_{N-1}(k+q_j;\,q_1+q_j,\ldots)
\label{WI2}
\eqe 
\[
+\,{1 \over A(\vk,q_j)}\,I_{N-1}(k;\,q_1+q_j,\ldots)-\,
{1 \over A(\vk+\vq_j,q_j)}
\,I_{N-1}(k+q_j;\,q_1+q_j,\ldots)\ .
\]
Here we used the definition
\eq
\De(\vk,q_j)\,=\,
{1 \over A(\vk+\vq_j,q_j)}\,-\,{1 \over A(\vk,q_j)}\,=\,
{2 \,\vq^{\,2}_j \over A(\vk+\vq_j,q_j) \, A(\vk,q_j)}\ .
\label{dede}
\eqe
Regarding (\ref{WI2}) we realize that the last two terms give a  
vanishing contribution on integration over $\,\vk\,$, by a shift
of $\,k\,$.  And the prefactors of the first two terms scale as $\la
^2\,$ in the dynamical limit (see below (\ref{ine}, \ref{boo})),
whereas there appears  a gain of a factor of $\la\,$ in the 
scaling limit when taking into account the change $\,N\to N-1\,$ 
in the second term and using an inductive argu\-ment based on
A2) (\ref{pcsy}). To make work this inductive argument, we     
have  to generalize (\ref{WI2}) to symmetrization
w.r.t. more than one variable.
The Ward identity for  
$I_N^{\si_N^n}\,$ with  $\,n\ge 2\,$ 
is obtained from (\ref{WIprop}) in the same way as
(\ref{WI})~: 
\eq
\sum_{\si_N(j_1,\ldots,j_n)}  
 \{iq_{j_{\nu},0}- 2 \vq_{j_{\nu}} \cdot
(\vk-\vp_{j_{\nu}}^{\,\si_N^n})+\vq^{\,2}_{j_{\nu}}\}\,   
I_N^{\si_N^n}(k;\,q_1,\ldots, q_{N})\,=\,
\label{WI3}
\eqe 
\[
\sum_{\si_{N-1}(j_1,\ldots,\not j_{\nu}, \ldots, j_n)}  
\biggl(I_{N-1}^{\si_{N-1}^{n-1}}
(k;\,q_1+q_{j_{\nu}},\ldots, q_{N})\,-\,
I_{N-1}^{\si_{N-1}^{n-1}}
(k+q_{j_{\nu}};\,q_1+q_{j_{\nu}},\ldots, q_{N}) \biggr)\ .
\]
Here the momentum  $\,\vp_{j_{\nu}}^{\,\si_N^n}\,$ is 
the one arriving at the vertex of $q_{j_{\nu}}\,$,
$\,j_{\nu}\in\{j_1,\ldots,j_n\}\,$,   for the per\-mutation
$\,\si_{N}^n\,$. For each permutation $\,\si_N^n\,$
appearing on the l.h.s. we sum on the r.h.s. over a permutation 
$\,[\si_{N-1}^{n-1}(\si_{N}^{n},j_{\nu})](j_1,\ldots,\not \! j_{\nu},
\ldots, j_n) \,$ which is defined
as
\eq
[\si_{N-1}^{n-1}(\si_{N}^{n},j_{\nu})](j)\,= 
\left \{\!\! \begin{array}{r@{,}l} 
\,\si_{N}^{n}(j) \,& \ \mbox{ if }
\quad \si_{N}^{n}(j) <  \si_{N}^{n}(j_{\nu}) \\  
\si_{N}^{n}(j) -1 & \ \mbox{ if }
\quad  \si_{N}^{n}(j) >  \si_{N}^{n}(j_{\nu})  \  \\
\end{array}  \right \} 
\eqe
(so that $\,\si_{N-1}^{n-1}(\si_{N}^{n},j_{\nu})\,$
is indeed a map onto  $\,\{2,\ldots, N-1\}\,$).
To proceed to an identity in terms of $I_N^{S_{n}}\,$ we have to 
analyse and eliminate (as far as possible) the dependence  
of the term $\sim \vq_{j_{\nu}} \cdot
\vp_{j_{\nu}}^{\,\si_{N}^n}\,$ on the permutations $\si_N^n\,$. We use the
following \\[.2cm]
{\bf Lemma 2~:}\\
\eq
a)\qquad \qquad\qquad\,\, \sum_{\nu=1}^n \vq_{j_{\nu}}
\cdot\vp_{j_{\nu}}^{\,\si(j_1,\ldots,j_n)} =
\sum_{i,j,i<j \atop i,j \in \{j_1,\ldots,j_n\}} \!\!\!\! \vq_i \cdot\vq_j
\;+\,
\sum_{\nu=1}^n \vq_{j_{\nu}}
\cdot\vec{{\hat p}}_{j_{\nu}}^{\,\si(j_1,\ldots,j_n)} \ .
\label{deco}\qquad \quad
\eqe
Here $\hat p\,$  is obtained from $p\,$ by setting to zero the momenta
$q_{j_1},\ldots,q_{j_n}\,$, i.e.
\eq
{\hat p}_{j_{\nu}}^{\,\si(j_1,\ldots,j_n)}(q_1,\ldots,q_N)=
p_{j_{\nu}}^{\,\si(j_1,\ldots,j_n)}({\hat q}_1, \ldots,{\hat q}_N)\,,
\eqe
 where 
$\,{\hat q}_i = q_i\,$, if $ \, i \not\in \{j_1,\ldots,j_n\}\;$  and 
$\,\,{\hat q}_i = 0\,$,  if $ \, i \in \{j_1,\ldots,j_n\}\,$.
\eq
b)\qquad\qquad\qquad\
 {\hat p}_{j_{\nu}}^{\,\si(j_1,\ldots,j_n)}(q_1,\ldots,q_N)=
{\hat p}^{\,\hat{\si}_{j_{\nu}}}\,:=\!\!
\sum_{k\in (\{ 2,\ldots,N\}-\{j_1,\ldots, 
\not j_{\nu},\ldots, j_n\}), \atop
k<\,\hat{\si}^{-1}_{j_{\nu}}(j_{\nu}) }\!\!
q_{\hat{\si}_{j_{\nu}}^{-1}(k)} \ .\ \,\,
\qquad\quad\
\eqe
Here $\,\hat{\si}_{j_{\nu}}\,$ is 
a permutation of the type $\,\si_{N-\!(n-1)}^1\,$, and it is
defined as the permutation
of the sequence 
$\,\Bigl( (2,\ldots,N)-(j_1,\ldots, 
\not\!\! j_{\nu},\ldots, j_n) \Bigr)\,$
which transfers $j_{\nu}\,$ to the same position relative 
to  
$\,\Bigl( (2,\ldots,N)-(j_1,\ldots,  j_n) \Bigr)\,$ as
$\,\si(j_1,\ldots,j_n)\,$ 
does.
\eq
c)
\qquad\qquad\qquad\qquad\,
\sum_{\si(j_1,\ldots,j_{n})}
I_N^{\si(j_1,\ldots,j_n)}=
\sum_{\hat{\si}_{j_{\nu}}} 
\bigl(\,I_N^{S_{n-1}(j_1,\ldots,\not
  j_{\nu},\ldots,j_{n})}\bigr)^{\hat{\si}_{j_{\nu}}}  \ .\qquad\qquad \qquad
\label{eq}
\eqe
{\sl Proof~:}
a) To extract all terms $\sim  \vq_i \cdot\vq_j\,,\ \,i,\,j \in
\{j_1,\ldots,j_n\}\,$ from  
$\,\sum_{\nu=1}^n \vq_{j_{\nu}}
\cdot\vp_{j_{\nu}}^{\,\si(j_1,\ldots,j_n)} \,$,  we go through 
the sum over  $j_{\nu}\,$ according to the order in which
$q_{j_{\nu}}\,$ appears in the $N$-loop for the permutation
$\si(j_1,\ldots,j_{n})\,$, starting from the last momentum.
We realize that we pick up exactly once each pair 
$\vq_i \cdot\vq_j\,,\ i,\,j \in \{j_1,\ldots,j_n\}\,$. Once these
terms have been extracted the remainder obviously takes the form  
from (\ref{deco}). \\[.1cm]
b) Since 
${\hat p}_{j_{\nu}}^{\,\si(j_1,\ldots,j_n)}(q_1,\ldots,q_N)\,$  
equals the sum of the momenta arriving at the vertex of
$q_{j_{\nu}}\,$ in the permutation $\si(j_1,\ldots,j_{n})\,$, 
with all $\,q_j,\ j \in\{j_1,\ldots,j_n\}\,$ set to zero, it is
equal to the sum over those momenta, which lie  in the complementary set
and arrive at the vertex of $\,q_{j_{\nu}}\,$. So it equals
 $\,{\hat p}^{\,\hat{\si}_{j_{\nu}}}\,$.\\[.1cm]
c) On the r.h.s. of (\ref{eq}), $\,I_N^{S_{n-1}}\,$,
which has been symmetrized w.r.t. $\,(j_1,\ldots,
\not\!\!j_{\nu},\ldots,  j_n)\,$,
depends only on
the sequence  $\,\Bigl( (2,\ldots,N)-(j_1,\ldots,
\not\!\!j_{\nu},\ldots,  j_n) \Bigr)\,$, which is acted upon by 
$\,{\hat\si}_{j_{\nu}}\,$.
The statement (\ref{eq}) then follows from the observation 
that summing over all possible orderings of $\,(j_1, \ldots, j_n)\,$
within $(2, \ldots, N)\,$, keeping the order of the remaining
variables fixed, can be achieved  by summing, for fixed ordering of
$\,\Bigl( (2,\ldots,N)-(j_1,\ldots,\not\!\!
j_{\nu},\ldots,  j_n)\Bigr)\,$, over all
possible orderings of $\,(j_1, \ldots,\not\!\! j_{\nu},\ldots, j_n)\,$
within $(2, \ldots, N)\,$, and then 
over the position of $\,j_{\nu}\,$ relative to 
$\,\Bigl( (2,\ldots,N)-(j_1,\ldots, j_n)\Bigr)\,$.\qed \\[.2cm] 
Using Lemma 2 we come back to the analysis of (\ref{WI3}). 
On summing over $\,\nu\,$ we obtain:
\eq
\biggl[ \sum_{\nu =1}^n  
 \{iq_{j_{\nu},0}- 2 \vq_{j_{\nu}} \cdot
\vk+ \vq^{\,2}_{j_{\nu}}\}\,+\sum_{i,j,i<j \atop i,j \in
  \{j_1,\ldots,j_n\}} \vq_i \cdot\vq_j \biggr]    
I_N^{S_{n}(j_1,\ldots,j_n)}(k;\,q_1,\ldots, q_{N})\,=\,
\label{wind}
\eqe
\[
-\,2\,\sum_{\nu=1}^n \sum_{\hat{\si}_{j_{\nu}}} \vq_{j_{\nu}}
\cdot\vec{{\hat p}}_{j_{\nu}}^{\,\hat{\si}_{j_{\nu}}}\,
\bigl(I_N^{S_{n-1}(j_1,\ldots,\not
  j_{\nu},\ldots,j_{n})}\bigr)^{\hat{\si}_{j_{\nu}}} 
(k;\,q_1,\ldots, q_{N})
\] 
\[
+\,\sum_{\nu=1}^n \sum_{\si_{N-1}^{n-1}}
\biggl( I_{N-1}^{\si_{N-1}^{n-1}}
(k;\,q_1+q_{j_{\nu}},\ldots,\not\! q_{j_{\nu}}, \ldots,  q_{N})\,-\, 
I_{N-1}^{\si_{N-1}^{n-1}}
(k+q_{j_{\nu}};\,q_1+q_{j_{\nu}},\ldots,\not\! q_{j_{\nu}}, \ldots,  q_{N})
\biggr)\ .
\] 
For the prefactor on the l.h.s. of (\ref{wind}) 
we write
\eq
A(\vk;\,q_{j_1},\ldots,q_{j_n}):=\sum_{\nu =1}^n  
 \{iq_{j_{\nu},0}- 2 \vq_{j_{\nu}} \cdot
\vk+ \vq^{\,2}_{j_{\nu}}\}\,+\sum_{i,j,i<j \atop i,j \in
  \{j_1,\ldots,j_n\}} \vq_i \cdot\vq_j\;, 
\label{pref}
\eqe
and we also introduce 
\eq
\De(\vk,\vq_{j_{\nu}};\,q_{j_1},\ldots,q_{j_n}):=
{1 \over A(\vk+\vq_{j_{\nu}};\,q_{j_1},\ldots)}
-\,{1 \over A(\vk;\,q_{j_1},\ldots)}
=\,{2\sum_{\mu =1}^n  \vq_{j_{\mu}}\cdot\vq_{j_{\nu}}  
  \over 
A(\vk;\,\ldots)\,
A(\vk+\vq_{j_{\nu}};\ldots)}\ .
\label{preff}
\eqe
We divide by $\,A(\vk;\,q_{j_1},\ldots,q_{j_n})\,$ 
(similarly as in (\ref{WI2}) above)  and obtain
\[
I_N^{S_{n}}=\,
{-2 \over A(\vk;\,q_{j_1},\ldots)}
\sum_{\nu=1}^n \sum_{\hat{\si}_{j_{\nu}}}
\vq_{j_{\nu}}\cdot\vec{{\hat p}}_{j_{\nu}}^{\,\hat{\si}_{j_{\nu}}}\,
\bigl(I_N^{S_{n-1}}\bigr)^{\hat{\si}_{j_{\nu}}}
\,+\,\sum_{\nu=1}^n \De(\vk,\vq_{j_{\nu}};\ldots)
\,I_{N-1}^{S_{n-1}}(k+q_{j_{\nu}};q_1+q_{j_{\nu}},\ldots)
\]
\eq
+\,\sum_{\nu=1}^n 
\biggl(
{1 \over A(\vk;\ldots)}\,I_{N-1}^{S_{n-1}}(k;q_1+q_{j_{\nu}},\ldots)\,-\,
{1 \over A(\vk+\vq_{j_{\nu}};\ldots)}\,
I_{N-1}^{S_{n-1}}(k+q_{j_{\nu}};q_1+q_{j_{\nu}},\ldots)
\biggr)\ .
\label{sywind}
\eqe
In the first line on the r.h.s. 
there appear the terms
$\,I_N^{S_{n-1}}\,$ and $\,I_{N-1}^{S_{n-1}}\,$.
Regarding their  prefactors,  
$\,\vq_j \cdot\vp^{\,\hat{\si}_{j_{\nu}}}\,$ scales
as $\,\la^2\;$ in the small $\,\la\,$ and dynamical limits,
and, by (\ref{nonex}),
\eq
|A(\vk;\,q_{j_1},\ldots,q_{j_n})|\,>\,\eta\,,\quad
|A(\vk;\,\la q_{j_1},\ldots,\la q_{j_n})|\,>\,\la\,\eta\,, \quad
|A(\vk;\,q_{j_1,0},\,\la \vq_{j_1},\ldots)|\,>\,\eta\ \,,  
\label{ine}
\eqe
\eq
|\De(\vk,\la\vq_{j_{\nu}};\la q_{j_1},\ldots, 
\la q_{j_{\nu}})|\,\le {K_1 \over \eta ^2}\;,\ \
\De(\vk,\la\vq_{j_{\nu}};q_{j_1,0},\la \vq_{j_1},
\ldots, q_{j_{\nu},0},\la \vq_{j_{\nu}})|\,\le \la^{2}\,
{K_2 \over \eta ^2}\ ,
\label{boo}
\eqe
(where $\,K_1$, $K_2\,$ depend on the (compact) sets of momenta considered).

Our inductive proof of A2) B2) is based on (\ref{sywind}) together
  with (\ref{int22}, \ref{int23}). 
We use an inductive scheme proceeding upwards
in $\,N\ge 3\,$, and for fixed  $\,N\,$ upwards
in $\,n\,$ for $0\le n\le N-2$.
The induction hypotheses are
\eq
|\Pi_N^{S_n}(\vph_s;\,\la q_1,\ldots, \la q_{N})|\le 
O(\la ^{-(N-2-n)})\,,\ \
|\Pi_N^{S_n}(\vph_d;\,q_{10},\la \vq_1,\ldots, \,q_{N0},\la \vq_N)|\le 
O(\la^{2+2n})\ .
\label{int24}
\eqe
For any $\,N\,$ and $\,n=0\,$ (the unsymmetrized case) the claim
follows from  (\ref{int22}). For $\,n>0\,$ 
we regard the scaling resp. dynamical limit for (\ref{sywind}),
multiplied by $\,\vph\,$ and integrated over $\,k\,$.
We can apply the induction
hypothesis to the r.h.s. of (\ref{sywind})  
noting again that the functions $\De\,\vph$ and $A\, \vph$ have the
properties reqired for $\vph\,$. We  also use (\ref{int23}) 
if $N=3\,$.  For each entry in the sum in the second line of
(\ref{sywind}) we perform in the second term the change of variables 
$\,\ti k =k+ \la q_j\,$ resp. $\,\ti k =k+ (q_{j0},\la \vq_j)\,$ and
  then use (\ref{bd}). With the aid of (\ref{ine}, \ref{boo}) 
and the induction hypothesis one then shows
\eq
|\int {d k_0 \over 2\pi}\,{d^dk \over (2\pi)^d}\,
I_N^{S_{n}}(k;\,\la q_1,\ldots, \la q_{N})
\,\vph(\vk;\,\la q_1,\ldots, \la q_{N})\,|
\,\le O(\la^{-(N-2-n)})\ , 
\label{fin1}
\eqe
\eq
|\int  {d k_0 \over 2\pi}\,{d^dk \over (2\pi)^d}\,
I_N^{S_{n}}(k;\,q_{1,0},\,\la\vq_1,\ldots, q_{N,0},\,\la\vq_{N})
 \,\vph(\vk;\,q_{1,0},\,\la\vq_1,\ldots, q_{N,0},\,\la\vq_{N})
\,|
\,\le O(\la^{2+2n})\ . 
\label{fin2}
\eqe
On specializing to $\,\vph\equiv 1\,$, 
this ends the proof of the proposition. \qed\
\newpage
\noindent
We join a few comments on various extensions of the results 
obtained.\\ 
a) For {\it dimensions $\,d\ge 4\,$ }
the $\,N$-loop integrals are absolutely convergent
for $\,2N>d+1\,$ and can be obtained as limits $\,\Lao\to\infty$
of their regularized versions, which are defined
on introducing a regulating function
$\,\rho({\vk^2 \over \Lao ^2})\,$ in the propagators
\[
G_0(k)\to\, G_0(\Lao,k)\,=\,
{1 \over ik_0\,-\,(\rho^{-1}({\vk^2 \over \Lao ^2})\, \vk^2 \,-\,1)}\ . 
\]
We suppose $\,\rho\,$ to be smooth, monotonic, 
positive, of fast decrease and such that 
$\,\rho(x)\equiv 1\,$ for $\,x\le 1\,$. The regulator then appears in the
$\,A\,$-factors when using the Ward identity, e.g. (\ref{ad}) changes
into 
\[
iq_{j,0}\,+\,\rho^{-1}({(\vk-\vq_j)^2  \over \Lao ^2})\, (\vk-\vq_j)^2 \,
-\,\rho^{-1}({\vk^2  \over \Lao ^2})\,\vk^2 \ .
\]
But since these factors are still independent of $\,k_0\,$, the
regulator disappears without leaving any trace after performing the  
$\,k_0\,$-integration, if $\Lao \ge \sum |\vq_j| +1 \,$. So we still
obtain the same results for $\,d\ge 4\,$, if $\,2N>d+1\,$, and we 
obtain them without this last restriction in case  we define the integrals 
as $\Lao\to \infty\,$- limits of their regulated versions from the
beginning.\\ 
b) Neumayr and Metzner [1] also prove
$\,|\Pi_N^S(q_1,\ldots,q_N)| \le O(|\vq_j|)\,$
for $\,\vq_j \to 0\,$, keeping the other variables fixed. 
In our framework this result is obtained immediately from
(\ref{WI2}), and we realize that it holds already on symmetrization
with respect to $\,\vq_j\,$, full symmetrization is not required. 
This result can be generalized 
to several vanishing external
momenta  $\,\vq_{j_1}\,,\ldots\,,\vq_{j_n}\, $, in the same way as we
did for the proof of A2) and B2) in the proposition. Using
(\ref{sywind}) we obtain on induction 
\eq
|\Pi_N^{S_n(j_1,\ldots,j_n)}(q_1,\ldots,q_N)| \le 
O(\prod_{\nu=1}^{n} |\vq_{j_{\nu}}|) 
\label{bsp}
\eqe
and of course the same bound on $\,\Pi_N^{S}\,$.\\ 
c) From the proof one can straightforwardly read off a  bound w.r.t. 
the dependence on the parameter $\,\eta\,$ from (\ref{nonex}).
This bound is in terms of $\,\eta^{-(N+n)}\,$, stemming from the
contributions with a maximal number of factors of $\,\De\,$.
It is of course rather crude, since it does not take into
account the effects of the nonvanishing spatial variables
and can be improved, depending on the hypotheses made on those.\\   

In conclusion we have recovered previous results on the infrared 
behaviour of the connected $N$-point 
density-correlation functions, in short  $N$-loops, 
by simple, but rigorous arguments based on the {\it Ward identity}. 
We obtain bounds for the fully symmetrized $N$-loop,
in showing, how successive symmetrization improves the infrared
behaviour.\footnote{ We recently learned from W. Metzner, 
that they were also aware of the fact that partial
symmetrization improves the infrared behaviour, 
but did not mention it in [1].}
The bounds hold in any spatial dimension 
(taking into account  the remarks from  a) above).
Since the Ward identities are explicit and easy to handle,
they permit generalizations such as (\ref{bsp}).\newpage
\noindent
{\bf Acknowledgement}~: We would like  to thank Walter Metzner
for  acquainting us with the results from [1] and comments on a
previous version of the paper. We are particularly indebted to 
Manfred Salmhofer for detecting a major mistake in
our first version, suggesting quite a number of
further corrections and improvements 
and for many  valuable comments on the paper.

\medskip
\noindent{\bf References}
\medskip

\noindent
[1] A. Neumayr and W. Metzner, Phys. Rev. {\bf B58}, 
15449 (1998), and J. Stat. Phys. {\bf 96}, 613 (1999). 

\noindent
[2] J. Feldman, H. Kn{\"o}rrer, R. Sinclair and E. Trubowitz, 
in {\sl Singularities}, edited by M. Greuel (Birkh{\"a}user, Basel, 1998). 

\noindent
[3] W. Metzner, C. Castellani and C. di Castro, Adv. Phys. {\bf 47}, 
317 (1998).

\noindent
[4] P. Kopietz, {\sl Bosonization of Interacting Fermions in Arbitrary
Dimensions}, (Springer, Berlin,  1997). 

\noindent
[5] P. Kopietz, J. Hermisson and K. Sch{\"o}nhammer, 
Phys. Rev. {\bf B52}, 10877 (1995).

\noindent
[6] F. Stern, Phys. Rev. Lett. {\bf 18}, 546 (1967).

\end{document}